\input{aipcheck.tex}

  \def\selectedoptions{final}

\documentclass[
   \selectedoptions 
  ]
  {aipproc}

\def\lsim{\mathrel{\rlap{\lower 4pt \hbox{\hskip 1pt $\sim$}}\raise 1pt \hbox
        {$<$}}}
\def\gsim{\mathrel{\rlap{\lower 4pt \hbox{\hskip 1pt $\sim$}}\raise 1pt \hbox
        {$>$}}}

\newcommand{\halflife}{$t_{1/2}$\ }

\layoutstyle{6x9}

\SetInternalRegister\hbadness{8000} 

\newcommand\doingARLO[2][]{%
  \ifx\mmref\undefined #1\else #2\fi
}

\begin{document}

{\footnotesize
Invited talk at the ``GAMMA2001'' Symposium, April 4-6, 2001, Baltimore.

To be published in ``Gamma-Ray Astronomy 2001'' (American Institute of 
Physics) 
}

\title 
      [Gamma-Ray Signatures of Supernovae and Hypernovae]
      {Gamma-Ray Signatures of Supernovae and Hypernovae}

\classification{43.35.Ei, 78.60.Mq}
\keywords{}

\author{Ken'ichi Nomoto}{
  address={Department of Astronomy and Research Center for 
      the Early Universe, School of Science, University of Tokyo,
      Bunkyo-ku, Tokyo 113-0033, JAPAN},
  email={nomoto@astron.s.u-tokyo.ac.jp}  
}

\author{Keiichi Maeda}{
  address={Department of Astronomy and Research Center for 
      the Early Universe, School of Science, University of Tokyo,
      Bunkyo-ku, Tokyo 113-0033, JAPAN},
  email={maeda@astron.s.u-tokyo.ac.jp}
}

\author{Yuko Mochizuki}{
  address={RIKEN (The Institute of Physical and Chemical Research), 
Hirosawa 2-1, Wako, \\
Saitama 351-0198, JAPAN},
  email={} 
}

\author{Shiomi Kumagai}{
  address={Department of Physics, College of Science and Technology,
Nihon University, \\
Kanda-Surugadai 1-8, Chiyoda-ku, Tokyo 101, JAPAN},
  email={} 
}

\author{Hideyuki Umeda}{
  address={Department of Astronomy and Research Center for 
      the Early Universe, School of Science, University of Tokyo,
      Bunkyo-ku, Tokyo 113-0033, JAPAN},
  email={umeda@astron.s.u-tokyo.ac.jp} 
}

\author{Takayoshi Nakamura}{
  address={Department of Astronomy and Research Center for 
      the Early Universe, School of Science, University of Tokyo,
      Bunkyo-ku, Tokyo 113-0033, JAPAN},
  email={nakamura@astron.s.u-tokyo.ac.jp} 
}

\author{Isao Tanihata}{
  address={RIKEN (The Institute of Physical and Chemical Research), 
Hirosawa 2-1, Wako, \\
Saitama 351-0198, JAPAN},
  email={} 
}

\copyrightyear  {}

\begin{abstract}

We review the characteristics of nucleosynthesis and radioactivities
in 'Hypernovae', i.e., supernovae with very large explosion energies
($ \gsim 10^{52} $ ergs) and their $\gamma$-ray line signatures.  We
also discuss the $^{44}$Ti line $\gamma$-rays from SN1987A and the
detectability with INTEGRAL.  Signatures of hypernova nucleosynthesis
are seen in the large [(Ti, Zn)/Fe] ratios in very metal poor stars.
Radioactivities in hypernovae compared to those of ordinary
core-collapse supernovae show the following characteristics: 1) The
complete Si burning region is more extended, so that the ejected mass
of $^{56}$Ni can be much larger.  2) Si-burning takes place in higher
entropy and more $\alpha$-rich environment.  Thus the $^{44}$Ti
abundance relative to $^{56}$Ni is much larger.  In aspherical
explosions, $^{44}$Ti is even more abundant and ejected with
velocities as high as $\sim$ 15,000 km s$^{-1}$, which could be
observed in $\gamma$-ray line profiles.  3) The abundance of $^{26}$Al
is not so sensitive to the explosion energy, while the $^{60}$Fe
abundance is enhanced by a factor of $\sim$ 3.

\end{abstract}

\date{\today}

\maketitle

\section{Introduction}

Massive stars in the range of 8 to $\sim$ 100$M_\odot$ undergo
core-collapse at the end of their evolution and become Type II and
Ib/c supernovae (SNe II and SNe Ib/c).  These SNe II and SNe Ib/c
release large explosion energies and eject explosive nucleosynthesis
products, thus being major sources of radioactive species.  Until
recently, we have considered supernovae with the explosion energies of
$E =$ 1 - 1.5 $\times$ 10$^{51}$ ergs.  These energies have been
estimated from the observations of nearby supernovae, such as SNe
1987A, 1993J, and 1994I.  Also the progenitors of these SNe are
estimated to be 13 - 20 $M_\odot$ stars (e.g., \cite{nomoto00}).

Recently, there have been a number of candidates for the gamma-ray
burst (GRB)/supernova (SN) connection (see \cite{nomoto00} for 
references), including GRB980425/SNIc 1998bw, GRB971115/SNIc 1997ef,
GRB970514/SNIIn 1997cy, GRB980910/SNIIn 1999E, GRB980326, and
GRB970228.  Among the SNe with a possible GRB counterpart, SNeIc
1998bw \cite{iwamoto98,woosley99} and 1997ef \cite{iwamoto00,mazzali00} 
are characterized by a very large
kinetic explosion energy, $E \gsim 10^{52}$ erg.  This is more than
one order of magnitude larger than in typical SNe, so that these
objects may be called "Hypernovae".  These SNe produced more $^{56}$Ni
than the average core collapse SN.  The masses of these hypernova
progenitors are estimated to be $M \gsim 25 M_\odot$.  These massive
stars are likely to form black holes, while less massive stars form
neutron stars (see, however, \cite{wheeler00}).

Regarding $\gamma$-ray signatures of such hypernovae, whether and how
the hypernovae actually induce gamma-ray bursts needs further study of
aspherical explosions (e.g., \cite{macfadyen99,khokhlov99}).  Another
$\gamma$-ray signatures, we discuss here, are the line $\gamma$-ray
emissions.  We review the characteristics of nucleosynthesis and
radioactivity in hypernovae and their $\gamma$-ray line signatures.
We also discuss the $^{44}$Ti line $\gamma$-rays from SN1987A and the
detectability with INTEGRAL.  For line $\gamma$-rays from Type Ia
supernovae, see Kumagai \& Nomoto \cite{kumagai97} for a review.

Before discussing $\gamma$-rays, we first point out that signatures of
hypernova nucleosynthesis are seen in the large [(Ti, Zn)/Fe] ratios
in very metal poor stars.

\section {Nucleosynthesis in Hypernova Explosions}

In core-collapse supernovae/hypernovae, stellar material undergoes
shock heating and subsequent explosive nucleosynthesis. Iron-peak
elements are produced in two distinct regions, which are characterized
by the peak temperature, $T_{\rm peak}$, of the shocked material.  For
$T_{\rm peak} > 5\times 10^9$K, material undergoes complete Si burning
whose products include Co, Zn, V, and some Cr after radioactive
decays.  For $4\times 10^9$K $<T_{\rm peak} < 5\times 10^9$K,
incomplete Si burning takes place and its after decay products include
Cr and Mn (e.g., \cite{hashimoto89,woosley95,thielemann96}). 

We note the following characteristics of nucleosynthesis with very
large explosion energies \cite{nomoto01}:

1) Both complete and incomplete Si-burning regions shift outward in
mass compared with normal supernovae, so that the mass ratio between
the complete and incomplete Si-burning regions is larger.  As a
result, higher energy explosions tend to produce larger [(Zn, Co)/Fe],
small [(Mn, Cr)/Fe], and larger [Fe/O].  The elements synthesized in
this region such as $^{56}$Ni, $^{59}$Cu, $^{63}$Zn, and $^{64}$Ge
(which decay into $^{56}$Co, $^{59}$Co, $^{63}$Cu, and $^{64}$Zn,
respectively) are ejected more abundantly than in normal supernovae.  

2) In the complete Si-burning region of hypernovae, elements produced
by $\alpha$-rich freezeout are enhanced because nucleosynthesis
proceeds at lower densities (i.e., higher entropy) and thus a larger
amount of $^{4}$He is left.  Hence, elements synthesized through
capturing of $\alpha$-particles, such as $^{44}$Ti, $^{48}$Cr, and
$^{64}$Ge (decaying into $^{44}$Ca, $^{48}$Ti, and $^{64}$Zn,
respectively) are more abundant.

3) Oxygen burning takes place in more extended, lower density regions
for the larger explosion energy.  Hence, O, C, Al are burned more
efficiently and their abundances in the ejecta are smaller, while a
larger amount of burning products such as Si, S, and Ar are
synthesized.  Therefore, hypernova nucleosynthesis is characterized by
large abundance ratios of [Si/O], [S/O], [Ti/O], and [Ca/O].

\section {Aspherical Explosions}

Nakamura et al. \cite{nakamura01a} 
and Mazzali et al. \cite{mazzali01} have identified some
signatures of asymmetic explosion in the late light curve and spectra
of SN 1998bw.  Maeda et al. \cite{maeda00} have examined the effect of
aspherical (jet-like) explosions on nucleosynthesis in hypernovae.
The progenitor model is the 16 $M_\odot$ He core of the 40 $M_\odot$
star and the explosion energy is $E$ = 1 $\times$ 10$^{52}$ ergs.

\begin{figure}
    \begin{tabular}{c}
       \includegraphics[height=.3\textheight]{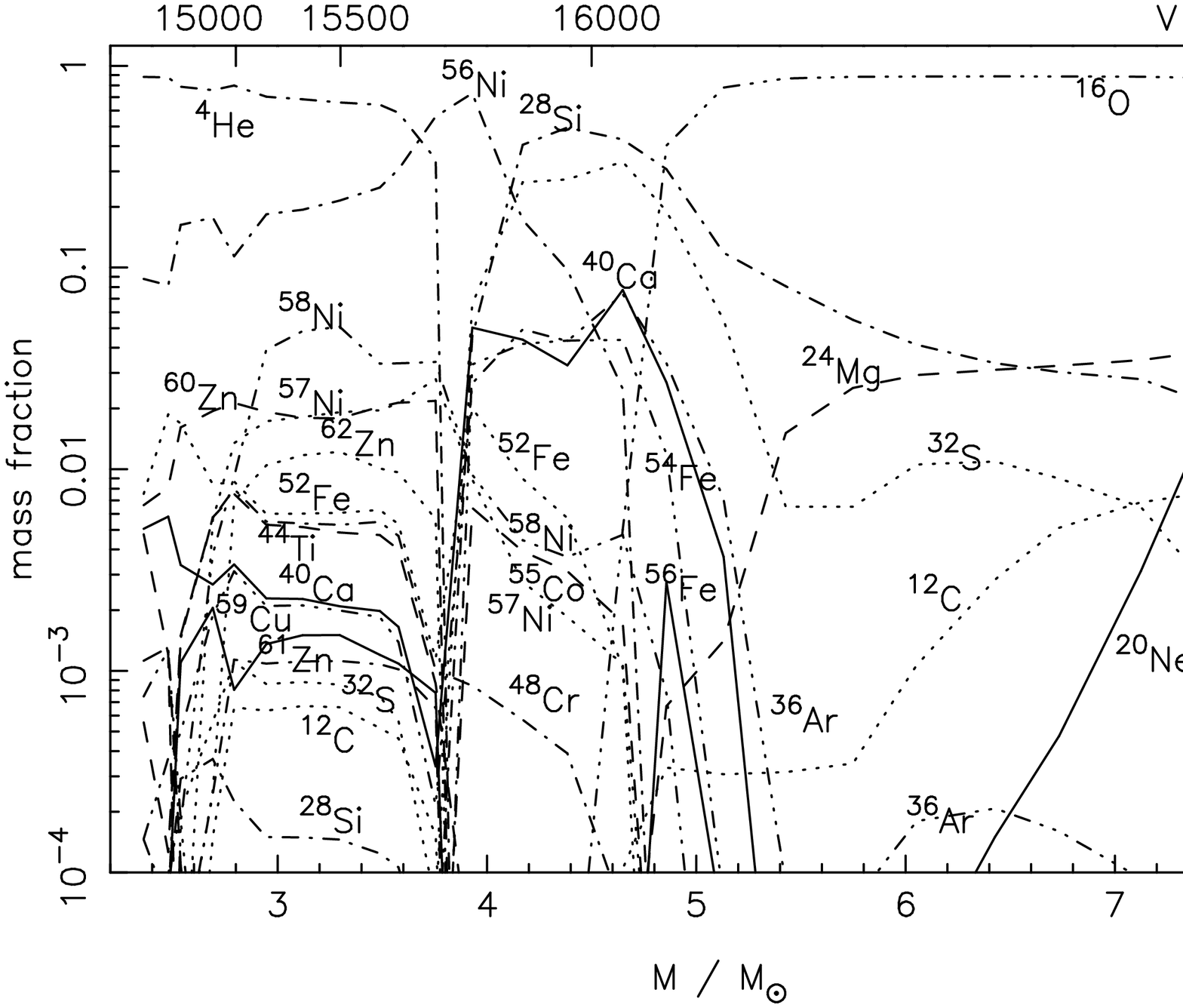}\\
       \includegraphics[height=.3\textheight]{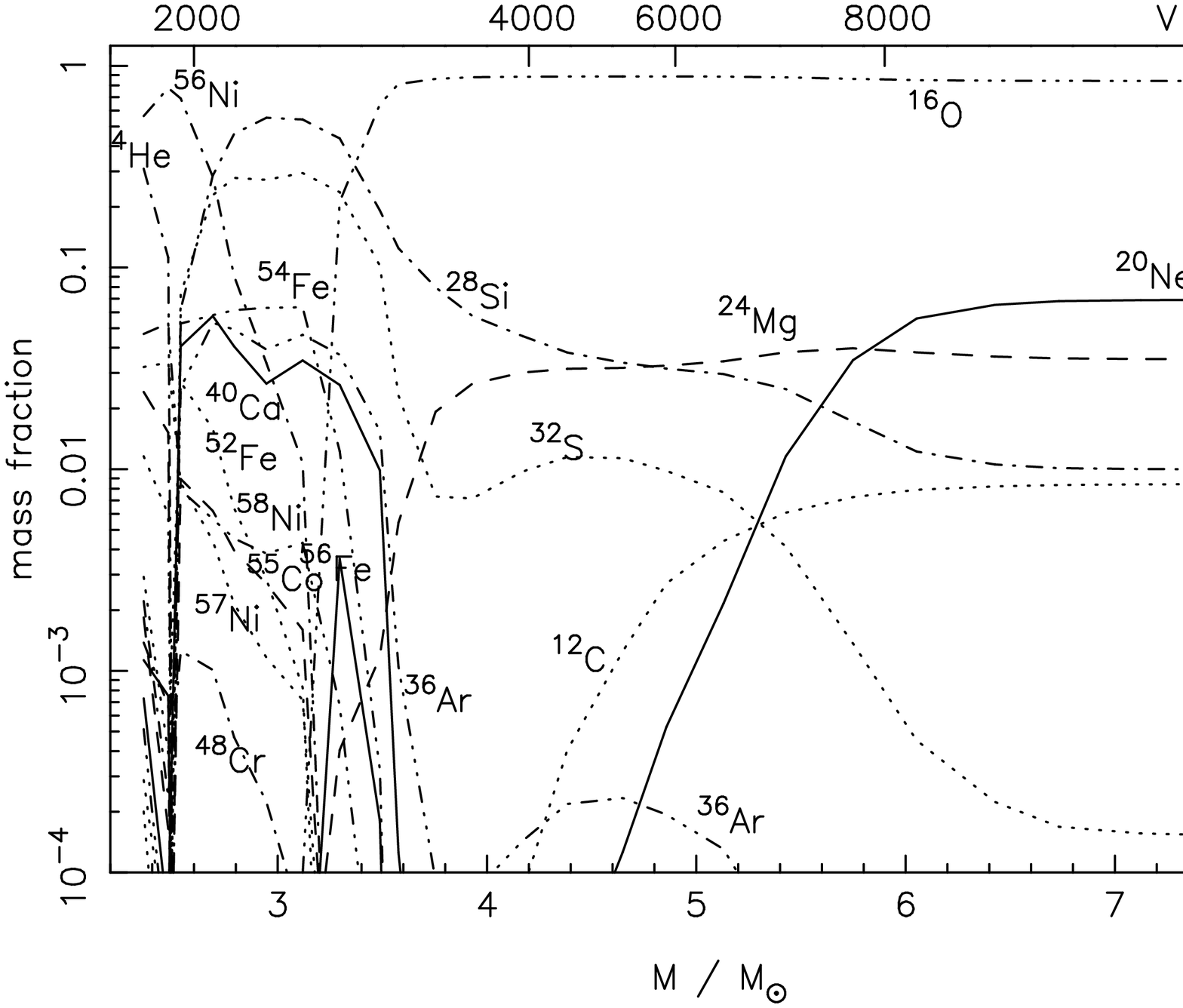}  
    \end{tabular}  
\caption{
The isotopic composition of the ejecta in the direction of 
the jet (upper panel) and perpendicular to it (lower panel).
The ordinate indicates the initial spherical Lagrangian coordinate ($M_r$) 
of the test particles (lower scale), and 
the final expansion velocities ($V$) of those particles (upper scale) 
\cite{maeda00}.
\label{fig:nuc1d}}
\end{figure}

Figure~\ref{fig:nuc1d} shows the isotopic composition of the ejecta of
asymmetric explosion model in the direction of the jet (upper panel)
and perpendicular to it (lower panel).  

In the $z$-direction, where the ejecta carry more kinetic energy, the
shock is stronger and post-shock temperatures are higher.  Therefore,
larger amounts of $\alpha$-rich freeze-out elements, such as $^4$He,
$^{44}$Ti, and $^{56}$Ni are produced in the $z$-direction than in the
$r$-direction.

On the other hand, along the $r$-direction $^{56}$Ni is produced only
in the deepest layers, and the elements ejected in this direction are
mostly the products of hydrostatic nuclear burning stages (O) with
some explosive oxygen-burning products (Si, S, etc). 

In the spherical case, Zn is produced only in the deepest layer, while
in the aspherical model, the complete silicon burning region is
elongated to the $z$ (jet) direction, so that [Zn/Fe] is enhanced
irrespective of the mass cut.  On the other hand, $^{55}$Mn, which is
produced by incomplete silicon burning, surrounds $^{56}$Fe and
located preferentially in the $r$-direction.

In this way, larger asphericity in the explosion leads to larger
[Zn/Fe] and [Co/Fe], but to smaller [Mn/Fe] and [Cr/Fe].  Then, if the
degree of the asphericity tends to be larger for lower [Fe/H], the
trends of [(Zn, Co, Mn, Cr)/Fe] follow the ones observed in metal-poor stars, 
as discussed later. 

\section{Signatures of Hypernova Nucleosynthesis in Galactic Chemical 
Evolution}

Several observational signatures of hypernova nucleosynthesis have
been noticed in several objects \cite{nomoto01}.
The abundance pattern of metal-poor stars with [Fe/H] $< -2$ provides
us with very important information on the formation, evolution, and
explosions of massive stars in the early evolution of the galaxy.

In the early galactic epoch when the galaxy is not yet chemically
well-mixed, [Fe/H] may well be determined by mostly a single SN event  
\cite{audouze95}. The formation of metal-poor stars is supposed
to be driven by a supernova shock, so that [Fe/H] is determined by the
ejected Fe mass and the amount of circumstellar hydrogen swept-up by
the shock wave \cite{ryan96}.  Then, hypernovae with larger $E$
are likely to induce the formation of stars with smaller [Fe/H],
because the mass of interstellar hydrogen swept up by a hypernova is
roughly proportional to $E$ \cite{ryan96,shigeyama98} 
and the ratio of the ejected iron mass to
$E$ is smaller for hypernovae than for canonical supernovae.

The observed abundances of metal-poor halo stars show quite
interesting pattern.  There are significant differences between the
abundance patterns in the iron-peak elements below and above [Fe/H]$
\sim -2.5$ - $-3$, which cannot be explained with the conventional 
chemical evolution model that uses previous nucleosynthesis yields.

1) For [Fe/H]$\lsim -2.5$, the mean values of [Cr/Fe] and [Mn/Fe]
decrease toward smaller metalicity, while [Co/Fe] increases  
\cite{mcwilliam95,ryan96}.

2) [Zn/Fe]$ \sim 0$ for [Fe/H] $\simeq -3$ to $0$ \cite{sneden91},
while at [Fe/H] $< -3.3$, [Zn/Fe] increases toward smaller metalicity 
\cite{primas00,blake01}.

The larger [(Zn, Co)/Fe] and smaller [(Mn, Cr)/Fe] in the supernova
ejecta can be realized if the mass ratio between the complete Si
burning region and the incomplete Si burning region is larger, or
equivalently if deep material from complete Si-burning region is
ejected by mixing or aspherical effects.  This can be realized if (1)
the mass cut between the ejecta and the collapsed star is located at
smaller $M_r$ \cite{nakamura99}, (2) $E$ is larger to move the outer
edge of the complete Si burning region to larger
$M_r$ \cite{nakamura01b}, or (3) asphericity in the explosion is
larger.

Also a large explosion energy $E$ results in the enhancement of the
local mass fractions of Zn and Co, while Cr and Mn are not enhanced 
\cite{umedanomoto01}.  Models with $E_{51} = E/10^{51}$ergs do not
produce sufficiently large [Zn/Fe].  To be compatible with the
observations of [Zn/Fe] $\sim 0.5$, the explosion energy must be much
larger, i.e., $E_{51} \gsim 20$ for $M \gsim 20 M_\odot$.  

Therefore, if hypernovae made significant contributions to the early
Galactic chemical evolution, it could explain the large Zn and Co
abundances and the small Mn and Cr abundances observed in very
metal-poor stars.

\begin{figure}
  \begin{minipage}[t]{0.5\textwidth}
    \includegraphics[height=.3\textheight]{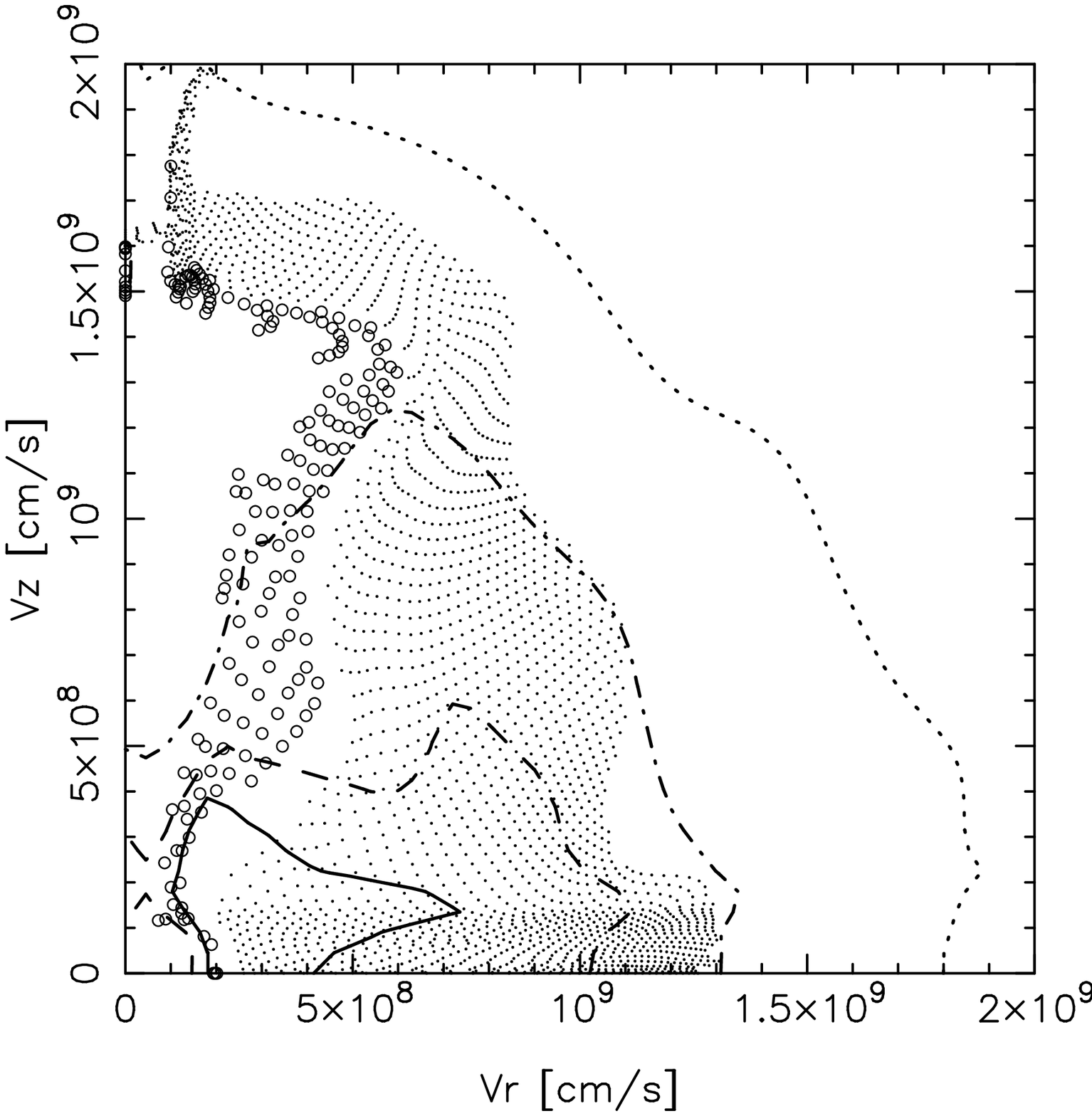}
  \end{minipage}
  \begin{minipage}[t]{0.5\textwidth}
    \includegraphics[height=.3\textheight]{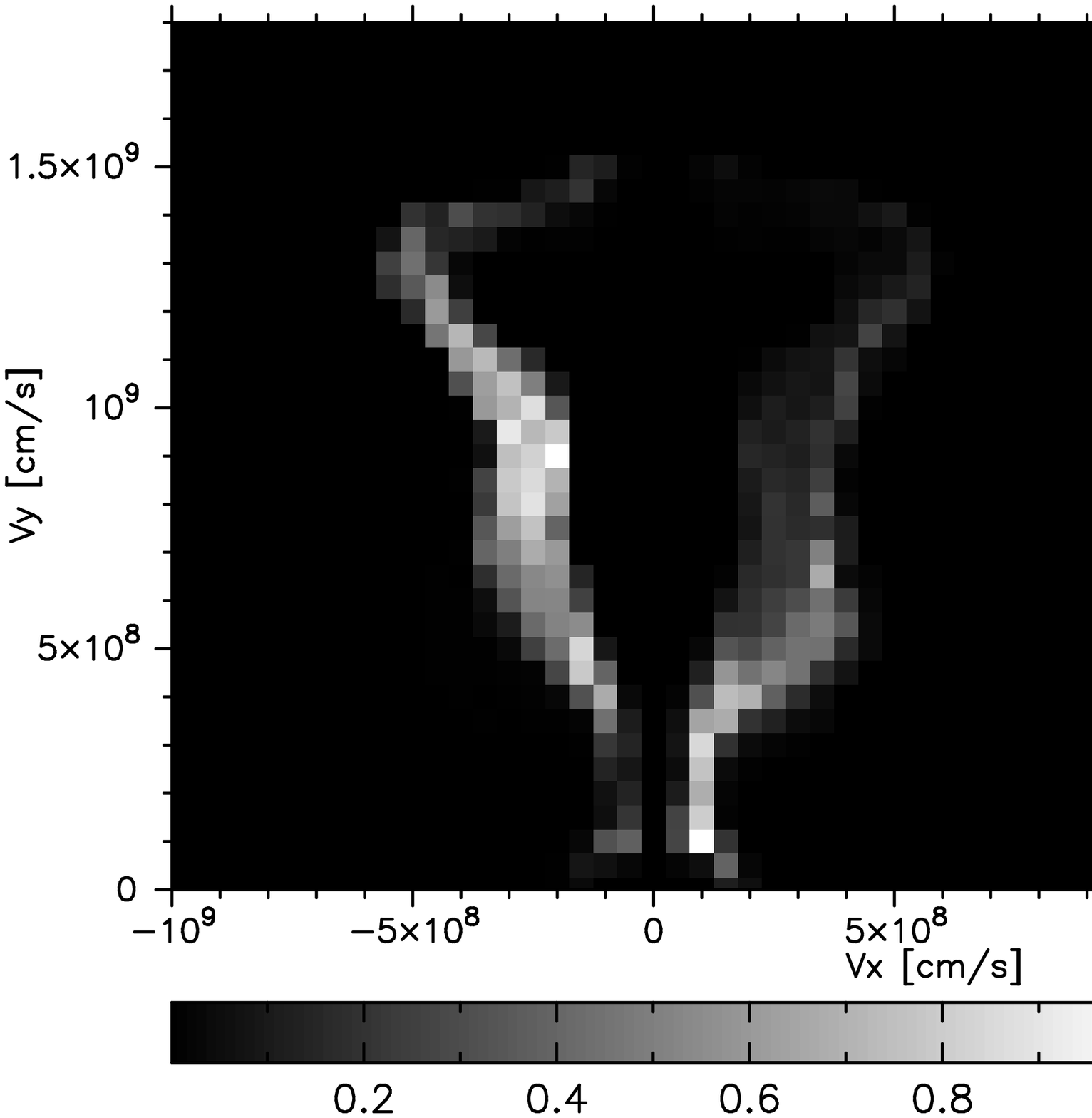}    
  \end{minipage}
\caption{
Left: The distribution of $^{56}$Ni (open circles) and $^{16}$O (dots) 
with density contours (lines) \cite{maeda00}.
Right: The density distributions of $^{44}$Ti (left half) and
$^{56}$Ni (right half). The densities of each element are
represented with linear scale, from 0 to the max density of
each element \cite{maeda01}.
\label{fig:2delm}}
\end{figure}

\section{Nuclear Radioactivity in Hypernovae}

Hypernovae produce radioactive species in larger amount than normal
SNe II, thus being important sources of line $\gamma$-rays and
positrons.

\subsection {$^{56}$Ni and Positrons}

A large amount of $^{56}$Ni can be produced as actually observed:
$\sim$ 0.5 $M_{\odot}$ from SN1998bw, $\sim 0.15 M_{\odot}$ from
SN1997ef, and even much larger from SN1999as.  Thus hypernovae are the
important source of positrons as well.  In particular, positrons in
the galactic center region might be significantly contributed by
hypernovae, because as observed in M82 \cite{nomoto01}, 
hypernovae would make an
important contribution in nucleosynthesis associated with starburst
events.

\subsection {$^{44}$Ti}

$^{44}$Ti is synthesized in the $\alpha$-rich freezeout, in complete
Si burning.  In the very high entropy environment in hypernovae, the
photons dissociate all the preexisting nuclei down to essentially
$\alpha$-particles and neutrons.  The abundances then shift to those
in nuclear statistical equilibrium but freezes out with excess
$\alpha$-particles.  The $^{44}$Ti yield depends strongly on the
location of the mass cut, electron fraction, and entropy condition in
the $\alpha$-rich freezeout.  Therefore the initial yield of $^{44}$Ti
provides unique information to constrain the explosion models.  

Iron-peak elements are produced in the deep region near the mass cut,
so that their production is strongly affected by asphericity of an
explosion.  Figure~\ref{fig:2delm} shows the 2D density distribution
of $^{44}$Ti and $^{56}$Ni, which is distributed preferentially in the
$z$-direction.  $^{44}$Ti, which is produced by the strong
$\alpha$-rich freezeout, is distributed preferentially in the
$z$-direction.  Moreover, $^{44}$Ti production is strongly enhanced
compared with a spherical model \cite{maeda01, nagataki97}, since the
post-shock temperature along the $z$-direction is much higher than
that of a spherical model.

Table~\ref{tab:radioactive} summarize the mass ratio
$^{44}$Ti/$^{56}$Ni produced in the spherical and the aspherical
explosion of the 40 $M_\odot$ models.  We can see that $^{44}$Ti is
strongly enhanced in hypernova and aspherical models \cite{maeda01}.  The
$\gamma$-rays from the decays of $^{44}$Ti are possible tools to
investigate the Galactic hypernova remnants.

\begin{table}
\begin{tabular}{ccc} \hline\hline
 Spherical ($E_{51} = 1$) & Spherical ($E_{51} = 10$) &
 Aspherical ($E_{51} = 10$) \\\hline
 6.51E-4 & 2.79E-3 & 1.01E-2 \\\hline
\end{tabular}
\caption{The mass ratio $^{44}$Ti/$^{56}$Ni for 40$M_{\odot}$ models 
\cite{maeda01}.
Values of spherical models are taken from Nakamura et al. \cite{nakamura01b}. 
\label{tab:radioactive}}
\end{table}

\subsection {$^{26}$Al and $^{60}$Fe}

Preliminary results of $^{26}$Al and $^{60}$Fe are summarized in
Table~\ref{tab:26al} \cite{umedanomoto01}.

1) The $^{26}$Al abundance does not depend much on $E$.  It is a
little smaller in hypernovae than supernovae, because more $^{26}$Al
is consumed in oxygen burning as are O, Mg, Ne.

2) $^{60}$Fe abundance is larger by a factor of $\sim$ 1.5 - 10.

\begin{table}
\begin{tabular}{ccccc} \hline\hline
  $E_{51}$   & 1      & 10     & 1      & 10     \\\hline
  $M$        &\multicolumn{2}{c}{$^{26}$Al} 
&\multicolumn{2}{c}{$^{60}$Fe}\\\hline
20$M_{\odot}$& 8.87e-5& 5.73e-5& 1.37E-5& 1.31E-4\\
25$M_{\odot}$& 1.17e-4& 1.22e-4& 1.86E-4& 3.38E-4\\
40$M_{\odot}$& 8.61e-5& 1.20e-4& 1.82E-5& 2.47E-5\\\hline
\end{tabular}
\caption{Mass of $^{26}$Al and $^{60}$Fe ($M_{\odot}$) (Z=0.02) 
\cite{umedanomoto01}. 
\label{tab:26al}}
\end{table}

\section {RX J0852-4622/GRO J0854-4622}

COMPTEL has detected the $^{44}$Ti 1157 keV $\gamma$-ray line 
from the supernova remnant (SNR) RX J0852-4622  
\cite{iyudin98,aschenbach98,aschenbach99}, though the
evidence is found at the 2$\sigma$ to 4$\sigma$ significance 
level \cite{schonfelder00}.  ASCA has observed RX J0852-4622 and
detected the 4.1 keV X-ray emission line from Ca \cite{tsunemi00}.  
Tsunemi et al. \cite{tsunemi00} have provided the following analysis
and interpretation.  The abundance of Ca is oversolar by a factor of 8
$\pm$ 5, while other elements are subsolar.  The mass of Ca is $\sim$
1.1 $\times$ 10$^{-3} M_\odot$.  The excess Ca is likely to be 
$^{44}$Ca, the decay product of $^{44}$Ti.  This feature is seen only
in the north-west shell, which suggests that the supernova ejecta has
just collided with the interstellar material there.

Iyudin \& Aschenbach \cite{iyudin01} have assumed that the width of the
$^{44}$Ti line is due to Doppler broadening and thus $^{44}$Ti is
expanding at $\sim$ 15,000 km s$^{-1}$ \cite{iyudin98}.  Then
they suggested that such high velocity $^{44}$Ti is ejected from a
Type Ia supernova of sub-Chandrasekhar mass \cite{livne95,arnett96}, 
because the model produces $\sim$ 10$^{-3} M_\odot$
$^{44}$Ti in the outer He detonation zone which expands at $\sim$
15,000 km s$^{-1}$.

Here we suggest an alternative model for the high velocity $^{44}$Ti.
As seen in Figure~\ref{fig:2delm}, the asymmetric hypernova explosion
ejects $^{44}$Ti at $\sim$ 15,000 km s$^{-1}$ in the jet direction.
The amount of $^{44}$Ti is as large as 1 $\times$ 10$^{-3} M_\odot$
(Table~\ref{tab:radioactive}), being consistent with the observation.

INTEGRAL observations of the $^{44}$Ti lines and their line profiles
from RX J0852-4622 are important to discriminates the models and
clarify the energetics of the explosion.

\section {SN~1987A}

SN 1987A in the LMC has shown for the first time that the energy
source of supernova ejecta directly comes from the decays of
radioactive nuclei (e.g., \cite{kumagai93,nomoto94b}) for
reviews).  In this section, we investigate the initial abundance of
$^{44}$Ti to discuss the detection possibility of the line gamma-rays
from SN~1987A, by comparing theoretical light curves with the observed
bolometric luminosity of SN~1987A.

In SN~1987A, it is established that the observed light curve in early
time is first governed by $^{56}$Ni [\halflife (half-life) = 6.1 d]
and then its daughter $^{56}$Co (\halflife = 77.3 d).  The synthesized
$^{56}$Co nuclide decays to stable
$^{56}$Fe.  As we shall see later, the observed light curve in the
wavelength ranging from ultraviolet (UV) to infrared (IR) has been
successfully modeled with the energy supply from the decay of
$^{56}$Co until $\sim$ 800 days (e.g., \cite{kumagai93}).  This has
been directly confirmed by the detection of the hard X-rays and the
line $\gamma$-rays from the decay sequence of $^{56}$Co.

Afterwards, the decline of the observed light curve apparently slowed
down, due to the decay of $^{57}$Co (\halflife = 272 d).
The slowness of the decline of observed light curve becomes
distinguished in particular after $\sim$ 1500 days from the explosion.
The dominant energy source at this moment can be attributed to
$^{44}$Ti decay.  $^{44}$Ti decays by orbital electron capture to
$^{44}$Sc, emitting 67.9 keV (100 \%) and 78.4 keV (98 \%) lines.
$^{44}$Sc then decays mainly by positron emission into $^{44}$Ca,
which emits a 1157 keV (100 \%) de-excitation line.

Obviously, a luminosity observation in SN~1987A of late years is
crucial to study the property of the extra energy source, namely, the
$^{44}$Ti production.  Recently, this crucial luminosity at 3600 days,
i.e., 10 years after the explosion was reported by Suntzeff 
\cite{suntzeff97}. 
The observed luminosity in the UV-IR range is
$L = (1.9 \pm 0.6)  \times 10^{36} \,  {\rm erg \, sec^{-1}}$.
The result is obtained by the collaboration of CTIO with HST, and
hereafter we refer to this as the CTIO+HST luminosity.
In the following, we investigate whether the $^{44}$Ti decay provides
enough energy to account for the observed CTIO+HST
luminosity under the latest half-life value of $^{44}$Ti 
\cite{mochizuki98}.

\subsection{The Half-Life of $^{44}$Ti}

The energy release from the $^{44}$Ti decay depends strongly on its
half-life.  
With recent experimental efforts,  
the half-life appears to be settled in 60 $\pm$ 3 
years, including the errors up to 3 $\sigma$ (see \cite{hashimoto01} 
and references therein).

We note here that the half-life values obtained in laboratories are
for neutral atoms.  Since $^{44}$Ti undergoes pure orbital electron
capture decay, its decay rate becomes smaller than the experimental
value if $^{44}$Ti is highly ionized under the condition of a
supernova.  For example, the decay rates of hydrogen-like and
helium-like ions are, respectively, $\sim$ 44 \% and $\sim$ 88 \% of
those of neutral atoms.  Details are found in 
Mochizuki et al. \cite{mochizuki99} and 
Mochizuki \cite{mochizuki01}.

It is expected that $^{44}$Ti was neutral when the CTIO+HST
observation was carried out.  We thus adopt 60 $\pm$ 3 years to
calculate theoretical light curves to compare the observation.

\subsection{Theoretical Light Curve of SN 1987A}

Our calculation of the light curves is based on Kumagai et al. 
\cite{kumagai93} 
and the adopted nuclear decay property data are updated.  We perform
Monte Carlo simulations of the Compton degradation of the line
$\gamma$-rays emitted from the decays of $^{57}$Co (14 keV, 122 keV,
136 keV, etc.) and $^{44}$Ti (68 keV, 78 keV, 1157 keV) to obtain the
UV-IR light curves.  The UV, optical, and IR photons originate from
the energy loss of the emitted $\gamma$-rays during the radiative
transfer in the ejecta.  The calculated UV-IR luminosity is the result
of subtracting the energy of the X-ray and $\gamma$-ray photons which
have managed to get out of the remnant.

For the velocity distribution of particles, we adopt the explosion
model 14E1 proposed by Shigeyama \& Nomoto \cite{shigeyama90} 
whose main-sequence
mass, ejecta mass, and explosion energy are 20 $M_\odot$, 14.6 $M_\odot$
(4.4 $M_\odot$ core material plus 10.2 $M_\odot$ hydrogen-rich
envelope), and $1 \times 10^{51}$ erg, respectively.  This model was
derived from a detailed analysis of the plateau shape of the light
curve of SN~1987A which is observed until 120 days after the
explosion, and well accounts for the earlier optical, X-ray, and
$\gamma$-ray light curves of SN 1987A \cite{nomoto91}.  Note that
the $^{56}$Ni mass in SN~1987A has been constrained as $0.07 M_\odot$
from the intensity during the observed exponential decline.  The
synthesized mass of the radioactive $^{57}$Ni is adopted from
Hashimoto, Nomoto, \& Shigeyama \cite{hashimoto89}, 
and the input value of the initial $^{44}$Ti
abundance is varied within the uncertainty studied in 
Hashimoto et al. \cite{hashimoto89} 
to compare the CTIO+HST luminosity.

\begin{figure}
  \includegraphics[height=0.5\textheight,angle=90]{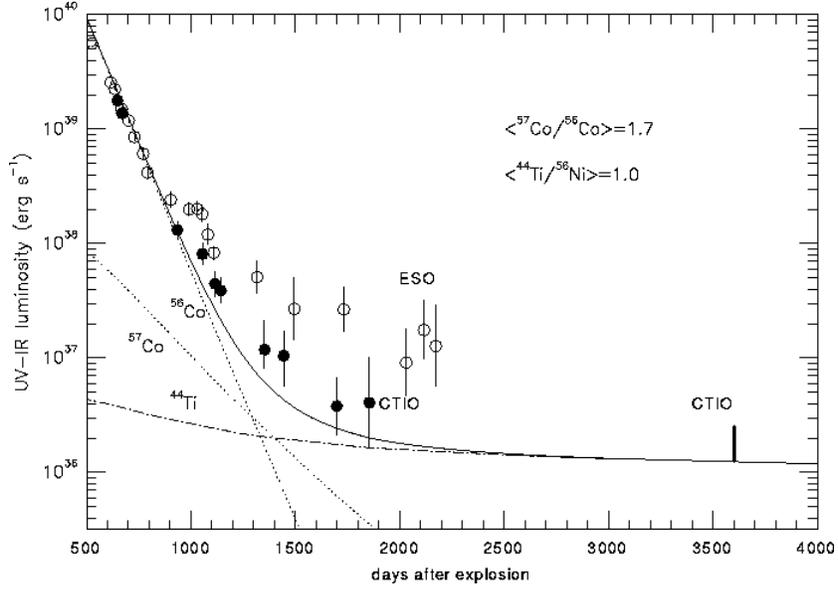}
\caption{The calculated light curve and the observed
bolometric (UV to IR) luminosity of SN~1987A, including the latest observed 
luminosity (CTIO+HST).
\label{fig:87a3}}
\end{figure}

In Figure~\ref{fig:87a3}, we show the calculated UV-IR light curve 
(solid) and the observed luminosities of SN1987A (CTIO and ESO). 
For this, the $^{44}$Ti 
half-life of 60 yrs and $<^{44}$Ti/$^{56}$Ni$> = 1$ have been adopted.
Here, $<^{44}$Ti/$^{56}$Ni$>$ is defined as the ratio of
$^{44}$Ti/$^{56}$Ni in SN1987A to 
$^{44}$Ca/$^{56}$Fe in the solar neighborhood, i.e.,
$<^{44}$Ti$/^{56}$Ni$> \equiv [X(^{44}{\rm Ti}) / X(^{56}{\rm Ni})] /
[X(^{44}{\rm Ca}) / X(^{56}{\rm Fe})]_\odot$.  The three decay
sequences of $^{56}$Ni, $^{57}$Ni and $^{44}$Ti are used for the energy
sources as shown in Figure~\ref{fig:87a3}. 
We found that the observed CTIO+HST luminosity
is reasonably explained with the energy release from the $^{44}$Ti decay
for its half-life of 60 $\pm$ 3 yrs.

\subsection{Production of $^{44}$Ti in SN~1987A}

The \mbox{$<^{44}$Ti/$^{56}$Ni$>$} values that are required to explain
the CTIO+HST luminosity are found to be roughly between 1 and 2.
Since the synthesized mass of $^{56}$Ni has been determined 
to be $0.07 M_\odot$  for  
SN~1987A, the obtained \mbox{$<^{44}$Ti/$^{56}$Ni$>$} values can
directly be translated to the initial mass of $^{44}$Ti.  Note that
the amount of $^{44}$Ti is the total value that is responsible for the
overall energy in $\gamma$-rays, $X$-rays, and in the UV-IR range.

We thus obtained that the initial $^{44}$Ti mass is $(1.1 - 2.5)
\times 10^{-4}$ $M_\odot$ for the half-life value of 60 $\pm$ 3 yrs.
We remark that the $^{44}$Ti yield calculated from recent explosive
nucleosynthesis models overlap with our estimate (e.g.,
\cite{thielemann90,timmes96,woosley91}).  However, it should be also
mentioned that these model predictions are subject to nuclear reaction
cross sections which have not yet measured so far (see, 
\cite{the98,sonzogni00}).

Note that Chugai et al. \cite{chugai97} 
estimated that positrons from (1-2)
$\times$ 10$^{-4} M_\odot$ of $^{44}$Ti provides the overall
luminosity of the FeII emission lines, and 
Lundqvist et al. \cite{lundqvist99,lundqvist01} 
obtained the upper limit of the $^{44}$Ti mass, 1 - 1.5 $\times
10^{-4}$ $M_\odot$, based on ISO SWS/LWS observations.

\begin{figure}
  \includegraphics[height=.35\textheight]{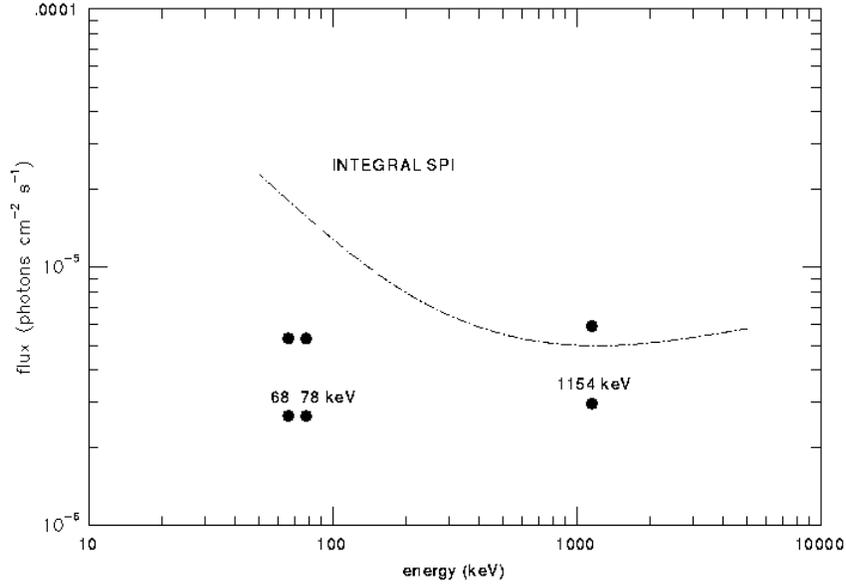}
\caption{Predictions of the line gamma-ray fluxes from the
$^{44}$Ti decay sequence (filled circles) for the initial mass of
$^{44}$Ti of 1 $\times$ 10$^{-4} M_\odot$ and 2 $\times$ 10$^{-4}
M_\odot$, respectively.  The sensitivity limit of INTEGRAL SPI (10$^6$
sec) is also shown by the dash-dotted curve.
\label{fig:87a6}}
\end{figure}

\subsection{$^{44}$Ti in SN~1987A with INTEGRAL}

Detection of $^{44}$Ti line $\gamma$-rays from SN~1987A should be an
important target for the INTEGRAL mission.  In
Figure~\ref{fig:87a6}, we show the expected fluxes from SN~1987A on
Earth for 68, 78, and 1157 keV lines, respectively.  For each
$\gamma$-ray line in Figure~\ref{fig:87a6}, the upper and lower 
filled circles are 
calculated for the $^{44}$Ti yield of 2 $\times 10^{-4}$ $M_\odot$,
and 1 $\times 10^{-4}$ $M_\odot$, respectively.  In
Figure~\ref{fig:87a6}, the sensitivity limit of INTEGRAL SPI (10$^6$
sec) is also shown, which strongly depends on energy.  We see from
Figure~\ref{fig:87a6} that it is possible to marginally detect the 
1157 keV line if the produced amount of $^{44}$Ti is as large as 2
$\times 10^{-4}$ $M_\odot$, but for other cases the expected fluxes
lie below the sensitivity limit of SPI.

In the above discussion of the
expected fluxes, no ionization effect on the $^{44}$Ti decay is taken
into account.  Recently, helium-like and hydrogen-like ions of O, Ne,
Mg, and Si have been observed with Chandra X-ray observatory 
\cite{burrows00}.  
The ionization of the ejected material was caused
by the shock heating associated with the collision with the 
circumstellar matter (including the ring).  As discussed previously, the
decay rate of highly ionized $^{44}$Ti can become considerably small
compared with that of neutral $^{44}$Ti.  This means, as claimed by
Mochizuki \cite{mochizuki01}, 
all the $^{44}$Ti $\gamma$-ray line fluxes from
SN~1987A possibly lie below the detection limit of INTEGRAL SPI.

Finally, we should note that INTEGRAL has capabilities both in the
detection of nuclear lines and pulsed emission.  It has not been clear
yet whether SN1987A has formed a neutron star or a black hole.  It is
certainly worth trying fo INTEGRAL and Astro E-II to search for
pulsation.

\begin{theacknowledgments}
 This work has been supported in part by the grant-in-Aid for
Scientific Research (07CE2002, 12640233) of the Ministry of Education,
Science, Culture, and Sports in Japan.
\end{theacknowledgments}

\end{document}